\documentclass[aps,amssymb]{revtex4}
\usepackage[dvips]{graphicx}

\begin{document}

\title{Complementary algorithms for graphs and percolation}

\author{Michael J.\ Lee}
\affiliation{Department of Physics and Astronomy, University of Canterbury, Christchurch, New Zealand}

\begin{abstract}
A pair of complementary algorithms are presented.
One of the pair is a fast method for connecting graphs with an edge.
The other is a fast method for removing edges from a graph.
Both algorithms employ the same tree based graph representation and so, in concert, can arbitrarily modify any graph.
Since the clusters of a percolation model may be described as simple connected graphs, an efficient Monte Carlo scheme can be constructed that uses the algorithms to sweep the occupation probability back and forth between two turning points.
This approach concentrates computational sampling time within a region of interest.
A high precision value of $p_\mathrm{c} = 0.59274603(9)$ was thus obtained, by Mersenne twister, for the two dimensional square site percolation threshold.
\end{abstract}

\maketitle
The various percolation models are well studied topological problems of statistical physics.
Beyond an intrinsic fundamental mathematical importance, percolation theory boasts a number of diverse applications in areas such as environmental science, biology, geology, chemistry, engineering, physics and cosmology \cite{Stauffer,Gould,Fairall}.

Contemporary Monte Carlo studies of percolation typically require that large numbers of samples be taken from within the critical region of the phase transition.
In order to take these samples a method such as that of Gould and Tobochnik \cite{Gould}, Machta, Choi, Lucke, Schweizer and Chayes \cite{Machta,MachtaPRE}, or Newman and Ziff \cite{Newman,NewmanPRE} is often employed.
These schemes all begin with an empty lattice and proceed to occupy individual sites (or bonds), one by one, until a spanning cluster exists.
A high rate of sampling is achievable but the majority of the data thus acquired is sourced from lattice configurations that lie outside of the critical region.
Such undesireable sampling is manifest as a serious computational inefficiency that would be better avoided.

Here a method is introduced that allows for individual sites (or bonds) to be switched back and forth between the occupied and unoccupied states.
This enables the lattice to take a random walk through configurations that dwell entirely within the region of interest.
Consequently a higher proportion (potentially all) of the sampled data usefully contributes to the final result.

This method relies upon two efficient algorithms for performing operations upon graphs.
The first algorithm adds edges to graphs and graphs together.
The second algorithm removes edges from graphs and splits graphs apart.
The scope of these algorithms is much more general than merely percolation.
The two algorithms share a common tree based representation for graphs which enables them to work efficiently with the same data structures; easily inserting and deleting both edges and vertices.
While they are both applicable to general directed graphs, only simply connected undirected graphs shall be considered here.

\begin{figure}
\includegraphics{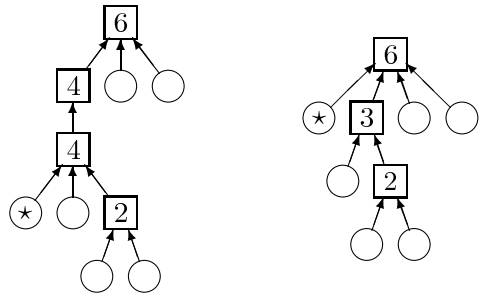}
\caption{The left hand tree of graph objects $\square$ and vertex objects $\bigcirc$ represents a graph. Data upon each graph object tracks the total number of vertex objects below. From the marked $\star$ vertex object the root is found by a recursive algorithm which follows pointers to the root and turns every object along the way into a child of the root. The result is the right hand tree which is a simpler representation of the same graph. The graph object that was a child of the root in the left hand structure has been pruned from the tree.}
\label{fig.tracing}
\end{figure}

Connected graphs are represented by trees of vertex objects and graph objects (see Fig.~\ref{fig.tracing}).
Vertex objects represent vertices, graph objects serve as a dynamic Hoshen-Kopelman relabelling table \cite{Hoshen,Stauffer} for the vertex objects and also maintain statistics (such as order) for the graph.
Each vertex object contains a pointer to the graph object of which it is a child.
Likewise each graph object contains a pointer to some other graph object of which it is a child.
Trees are constructed in such a way that for each tree there always exists one, and only one, graph object that is not the child of any other object and that has no pointer.
This unique object is designated as the root of the tree.
Vertex objects are always leaves of the tree, graph objects never are.
All vertex objects within a given tree are members of the same graph.
However it is not true, other than for the root, that the set of all vertex objects belonging to the sub-tree below a given graph object correspond to the vertices of a connected sub-graph.

The simplest possible graph consists of a single vertex and no edges.
To represent such a graph, a graph object is created sans pointer and a vertex object is created with its pointer directed at the graph object.
More complex graphs arise when the vertices of these simplest graphs become connected by edges.
Vertex objects carry (additional) pointers to any other vertex objects to which they are adjacent, and these pointers represent directed edges.
Undirected edges are represented by pairs of directed edges.
Edges are inserted into (or removed from) a graph simply by creating (or destroying) the corresponding pointers.

When two vertices from two different graphs become connected by an edge the two graphs must be fused into a single graph.
This is achieved by creating a pointer to the root of the greater order graph, upon the root of the lesser order graph, thereby making the lesser graph's root a child of the greater graph's root and the lesser graph's tree a sub-tree of the greater graph's tree.
Statistics upon the one remaining root are incremented by those on the former root so that the data upon the remaining root correctly reflects the properties of the new graph.
The result is a single tree with a unique root.

In order to perform this operation it is necessary to locate the root of each graph tree.
For every vertex object within the tree there exists a sequence of graph objects ending with the root.
For every object within the sequence the pointer upon that object uniquely specifies the next object in the sequence.
Starting with a vertex object the sequence is found by tracing pointers from child to parent until the root is reached.

A path compression technique is applied throughout the tracing process by redirecting the pointer of every object in the sequence (other than the root) to the location of the root.
In this way all objects in the sequence are made children of the root and future tracing operations from those objects will find the root after following only a single pointer.
This affords a significant improvement in computational efficiency \cite{Newman,NewmanPRE}.
To maintain correct statistics upon each graph object, relevant information is transferred from child to parent on every step of the tracing.
After path compression some graph objects from the sequence may have no children.
These redundant objects are pruned from the tree and deleted from memory during the tracing process.
The complete root finding process is shown in Fig.~\ref{fig.tracing}.

An edge may be removed from between two vertices by deleting the appropriate pointers from the corresponding vertex objects.
A vertex may be removed from a graph by removing all of the edges about that vertex (the associated vertex object may then be deleted).
Such operations may cause the original graph to fragment into a set of smaller graphs.
The deterministic accretion algorithm which efficiently identifies these fragments is similar to the stochastic methods of Hammersley and Handscomb \cite{Hammersley}, Leath \cite{Leath} and Alexandrowicz \cite{Alexandrowicz}, where sites and bonds are added to the perimeter of an existing percolation cluster.

Each surviving vertex that has had one or more of its edges removed is assigned to a distinct clump.
A clump is merely a label held in common by a set of vertices that are all known to belong to the same fragment graph.
Each such labelled vertex forms a separate nucleation kernel for an accretion process that constructs the fragment graphs from the remains of the original.
From each labelled vertex all edges are followed outward to what must be an adjacent vertex and these are assigned to the same clump as the labelled vertex.
As vertices are added to the perimeter of a clump, they are placed in a queue to be later examined for adjacent vertices which also need to be added to the clump.
This breadth first process is performed in parallel for all clumps and was found to be significantly faster than depth first accretion.

When two adjacent vertices are found to belong to different clumps, those two clumps are merged into one.
When a clump contains no vertex that is adjacent to any other vertex that is not a member of the same clump, then that clump has stopped growing and must represent a complete connected fragment graph. 
All vertex objects associated with the clump are then extracted from the original graph tree and are made children of a newly created (root) graph object.
Parameters for the fragment graph are calculated upon the root object and the clump ceases to be.

\begin{figure}
\includegraphics{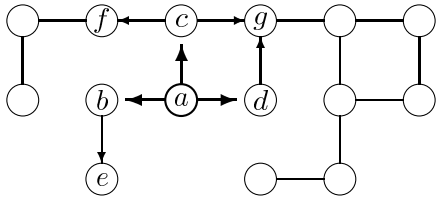}
\caption{Accretion algorithm. Nucleation: vertex $a$ and surrounding edges are removed; $b$, $c$ and $d$ are assigned to distinct clumps. Sweep one: $e$ assigned to clump of $b$; $f$ and $g$ assigned to clump of $c$; clumps of $g$ and $d$ merged. Sweep two: clump of $e$ complete and so is extracted to a graph; one clump remains hence all other vertices belong to the final fragment.}
\label{fig.accretion}
\end{figure}

When the number of distinct clumps remaining drops to one, whatever remains of the original graph tree must logically represent the final fragment graph.
It is not necessary to continue with the accretion process, the original graph tree is left as is, and the algorithm is complete.
As shown in Fig.~\ref{fig.accretion} the algorithm requires only enough time to establish the second largest fragment graph, rather than the largest, and the saving is often substantial.

Clump labelling is achieved by the use of clump objects to which vertex objects may or may not have pointers, exactly as in the graph trees.
The same pointer method is used to merge clumps as is used to fuse graphs.
The same path compressing pointer tracing process is used to find clump labels (clump tree root objects) as is used to find graph tree roots.
Use of clump objects permits the isolation of fragment graphs without altering the original graph tree data structure.
Since clumps are transient structures, statistics other than order need not be calculated for them.

It is possible to establish the fragment graphs without using clumps or clump objects.
Instead of giving vertices temporary clump labels, associated objects may be extracted directly into new graph trees.
These trees may later be fused back together during the accretion process.
However this requires additional work in extracting vertex objects from the original graph tree and in calculating graph statistics.
A significant performance difference was found in favour of using the clump approach.

Vertices can represent the binary state sites of a percolation model.
Graphs become equivalent to clusters of occupied sites simply connected by occupied bonds.
Unoccupied sites have neither graph object nor clump object pointers upon their corresponding vertex objects and are not members of any graph or clump.
An occupied site belongs to a cluster uniquely identified by the root of the graph tree to which the corresponding vertex object belongs.
Bonds may be occupied or unoccupied by switching edge pointers on and off upon the sites.
The algorithm presented here for joining two graphs becomes similar to the site-to-site pointer tree method of Newman and Ziff \cite{Newman,NewmanPRE}.
It is not entirely equivalent since it is impractical to do away with the graph objects while still retaining the edge and vertex removal component of the method.

Consider a site percolation model upon a fixed lattice of $N$ sites.
The pair of algorithms presented here confer the ability to arbitrarily raise and lower the number of occupied sites $n$ upon the lattice while efficiently maintaining the correct cluster information and associated statistics.
The algorithms may be forged into a Monte Carlo scheme that sweeps $n$ back and forth between two turning conditions.
Within this scheme, a Monte Carlo step consists of either the occupation of a randomly chosen unoccupied site, or the deoccupation of a randomly chosen occupied site.
Initially $n$ is stepped upward until the upper turning condition is satisfied and then $n$ is stepped downward until the lower turning condition is met.
This cyclical process may be repeated indefinitely.
This (bidirectional sweeping) approach is a generalisation of the single stopping condition (and implicit starting condition at $n=0$) found in the (unidirectional sweeping) algorithms of Machta, Choi, Lucke, Schweizer and Chayes \cite{Machta,MachtaPRE} and of Newman and Ziff \cite{Newman,NewmanPRE}.
While this method is conceptually simple, its computational performance depends strongly upon the details of how the Monte Carlo steps are realised.
The intention of the complementary graph algorithms described here is to achieve these steps in as little time as possible.

The turning conditions might be a change in the order parameter of the system.
The occupation $n$ can be increased until a spanning cluster exists and then decreased until a spanning cluster no longer remains.
The result is a self organised critical random walk through lattice configurations that are all only a single Monte Carlo step away from the phase change.
Such an approach hints at that of Tomita and Okabe \cite{Tomita}, where the current measured value of some order parameter determines a change in bond density.

In order to measure the square site percolation threshold $p_\mathrm{c}$ upon an $N = L \times L$ lattice, an unbiased Monte Carlo estimator of the spanning probabilities $R_L(p)$ \cite{Ziff,Reynolds} is useful.
The occupation probability $p = n/N$.
To provide this unbiased estimate the turning conditions are taken to be two fixed values of $n$ so that all configurations of a given $n$ are visited with equal probability.
Since knowledge of $R_L(p)$ is only required about the critical point, the turning points are chosen such that the sampling range includes this region.

Here lies an advantage over earlier unidirectional sweeping algorithms which return information for all $n$ from nought to the phase transition, a range of $O(N)$ steps.
The critical region spans a smaller range of only $O(NL^{-1/\nu})$ steps \cite{Stauffer}.
In two dimensions the correlation length exponent $\nu = 4/3$ \cite{Stauffer,Gould} and so the range is $O(N^{5/8})$.
Restricting the sampling range to the critical region significantly reduces the fraction of samples that make no contribution to the final result.
While the method presented here does not require any \emph{a priori} knowledge, it is able to make use of any information that does exist.
The performance and sampling range may be adjusted accordingly.

Estimating $R_L(p)$ requires knowing of the existence, or otherwise, of lattice spanning clusters.
An efficient means for tracking this information is integrated into the tracing process of Fig.\ \ref{fig.tracing}.
Each vertex object carries data as to upon which boundaries of the lattice the associated site lies.
When a site becomes occupied, this data is transferred to the root graph object where a running total is kept for all sites within the associated cluster.
It is then straightforward to determine which dimensions of the lattice are spanned by this cluster and that information is in turn sent from the root object to a master array for the entire lattice.
This array is updated whenever a cluster is modified so that at any given instant the precise number of clusters spanning any given combination of lattice dimensions is known.

Define $s_d(n)$ as the number of occasions on which the lattice is found to be in a configuration at occupation level $n$ within which there exists a cluster that spans the lattice across $d$, or more, dimensions.
Consequently $s_0(n)$ is the total number of observations made at occupation level $n$ and, in two dimensions, $s_2(n)$ is the number of those observations in which a single cluster spans across the entire lattice in both dimensions.
The number of observations in which a cluster spans either one, and only one, spatial dimension is given by $s_1(n) - s_2(n)$.
After making numerous observations, the probability that the lattice will be spanned over a given (specified) spatial dimension by a randomly generated configuration at occupation level $n$ is estimated by
\begin{displaymath}
R_L(n/N) = \frac{s_1(n) + s_2(n)}{2s_0(n)}
\end{displaymath}

It has been shown \cite{Ziff} that for large $L$ the spanning probability at the critical point is given by
\begin{displaymath}
R_L(p_\mathrm{c}) \approx 0.5 + b/L
\end{displaymath}
Ziff and Newman have found that $b = 0.320(1)$ \cite{ZiffPRE}.
Working on a lattice of $L = 2048$ ($N = 4194304$), it follows that $b/L = 0.00015625(50)$.

Calculations were performed to make estimates of $R_L(p)$.
Two sampling ranges were used; the entire critical region, $2474000 \leq n \leq 2498000$, and a small neighbourhood about the critical point, $2485700 \leq n \leq 2486700$.
In both cases the decorrelation time was found to be on the order of $5 \times 10^4$ steps.
This compares favourably with the $2.5 \times 10^6$ steps required to take a sample by unidirectional methods from an initially empty lattice.

Preliminary trials were conducted with a decimated Mitchell and Moore additive lagged Fibonacci generator with taps at 418 and 1279 \cite{Knuth}.
Results were consistent between using a single such generator and using an independently seeded pair to obtain $x$-$y$ site coordinates from only the most significant bits of each.
A combined total of $3 \times 10^9$ range sweeps yielded an estimate of $p_\mathrm{c} = 0.59274663(24)$.

\begin{figure}
\includegraphics{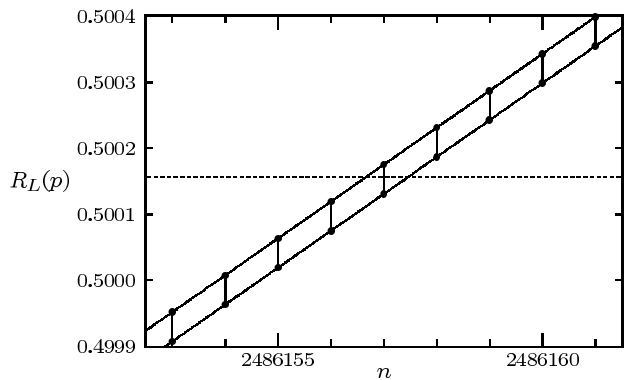}
\caption{Combined Mersenne twister based Monte Carlo estimates for the probability of the existence of a cluster spanning a single specified lattice dimension of a square site percolation model with $N = 2048^2$ sites. Solid lines are the error bounds for $R_L(p)$, the dashed line is $R_L(p_\mathrm{c})$.}
\label{fig.results}
\end{figure}

High precision measurements were conducted with Matsumoto and Nishimura's Mersenne twister generator MT19937 \cite{Matsumoto}.
Results were consistent between using a single generator and using a decimated independent $x$-$y$ pair.
A total of $2 \times 10^{10}$ range sweeps produced the estimates $R_L(2486156/N) = 0.500097(22)$, $R_L(2486157/N) = 0.500153(22)$ and $R_L(2486158/N) = 0.500209(22)$, as shown in Fig.~\ref{fig.results}.

It follows that the best estimate for the square site percolation threshold is $p_\mathrm{c} = 0.59274603(9)$, a measurement of significantly greater precision than earlier results \cite{Ziff,Newman,Deng} (see also Ref.~19 of \cite{Newman}) and which should assist with future studies of the model.
Note that this value is based purely upon the Mersenne twister calculations and differs from the lagged Fibonacci estimate.
At this level of precision, the choice of pseudorandom number generator is clearly of great importance.

In summary, efficient deterministic algorithms have been presented for the manipulation of graphs.
These are potentially useful in topological problems such as the analysis of networks and perturbative expansions in diagram formalisms.
It has been shown how they form the basis of an efficient Monte Carlo method that walks a system through every point within a chosen range and, crucially, only those points.
By exclusively sampling from pertinent system configurations, improved computational efficiency is achieved.
Greater quantities of useful data can be acquired than with other techniques and the result is an increased precision of measurement.
The method was used to achieve the most precise estimate to date of the square site percolation threshold and is applicable to a wide variety of numerical experiments on any discretised space of arbitrary connectivity.

Acknowledgements are due to  O.\ K.\ L.\ Petterson and C.\ J.\ McMurtrie for the provision of computing resources, and to R.\ M.\ Ziff for useful information and suggestions.
Elements of this work were conducted upon the University of Canterbury Supercomputer.

\bibliography{short}

\begin{thebibliography}{18}
\expandafter\ifx\csname natexlab\endcsname\relax\def\natexlab#1{#1}\fi
\expandafter\ifx\csname bibnamefont\endcsname\relax
  \def\bibnamefont#1{#1}\fi
\expandafter\ifx\csname bibfnamefont\endcsname\relax
  \def\bibfnamefont#1{#1}\fi
\expandafter\ifx\csname citenamefont\endcsname\relax
  \def\citenamefont#1{#1}\fi
\expandafter\ifx\csname url\endcsname\relax
  \def\url#1{\texttt{#1}}\fi
\expandafter\ifx\csname urlprefix\endcsname\relax\def\urlprefix{URL }\fi
\providecommand{\bibinfo}[2]{#2}
\providecommand{\eprint}[2][]{\url{#2}}

\bibitem[{\citenamefont{Stauffer and Aharony}(1994)}]{Stauffer}
\bibinfo{author}{\bibfnamefont{D.}~\bibnamefont{Stauffer}} \bibnamefont{and}
  \bibinfo{author}{\bibfnamefont{A.}~\bibnamefont{Aharony}},
  \emph{\bibinfo{title}{Introduction to Percolation Theory}}
  (\bibinfo{publisher}{Taylor and Francis}, \bibinfo{address}{London},
  \bibinfo{year}{1994}), \bibinfo{edition}{revised 2nd} ed.

\bibitem[{\citenamefont{Gould and Tobochnik}(1996)}]{Gould}
\bibinfo{author}{\bibfnamefont{H.}~\bibnamefont{Gould}} \bibnamefont{and}
  \bibinfo{author}{\bibfnamefont{J.}~\bibnamefont{Tobochnik}},
  \emph{\bibinfo{title}{An Introduction to Computer Simulation Methods}}
  (\bibinfo{publisher}{Addison-Wesley}, \bibinfo{address}{Reading, MA},
  \bibinfo{year}{1996}), \bibinfo{edition}{2nd} ed.

\bibitem[{\citenamefont{Fairall and Woudt}(2005)}]{Fairall}
\bibinfo{editor}{\bibfnamefont{A.~P.} \bibnamefont{Fairall}} \bibnamefont{and}
  \bibinfo{editor}{\bibfnamefont{P.~A.} \bibnamefont{Woudt}}, eds.,
  \emph{\bibinfo{title}{Nearby Large Scale Structures and the Zone of
  Avoidance}} (\bibinfo{publisher}{Astronomical Society of the Pacific},
  \bibinfo{address}{Provo, Utah}, \bibinfo{year}{2005}).

\bibitem[{\citenamefont{Machta et~al.}(1995)\citenamefont{Machta, Choi, Lucke,
  Schweizer, and Chayes}}]{Machta}
\bibinfo{author}{\bibfnamefont{J.}~\bibnamefont{Machta}},
  \bibinfo{author}{\bibfnamefont{Y.~S.} \bibnamefont{Choi}},
  \bibinfo{author}{\bibfnamefont{A.}~\bibnamefont{Lucke}},
  \bibinfo{author}{\bibfnamefont{T.}~\bibnamefont{Schweizer}},
  \bibnamefont{and} \bibinfo{author}{\bibfnamefont{L.~V.}
  \bibnamefont{Chayes}}, \bibinfo{journal}{Phys.\ Rev.\ Lett.}
  \textbf{\bibinfo{volume}{75}}, \bibinfo{pages}{2792} (\bibinfo{year}{1995}).

\bibitem[{\citenamefont{Machta et~al.}(1996)\citenamefont{Machta, Choi, Lucke,
  Schweizer, and Chayes}}]{MachtaPRE}
\bibinfo{author}{\bibfnamefont{J.}~\bibnamefont{Machta}},
  \bibinfo{author}{\bibfnamefont{Y.~S.} \bibnamefont{Choi}},
  \bibinfo{author}{\bibfnamefont{A.}~\bibnamefont{Lucke}},
  \bibinfo{author}{\bibfnamefont{T.}~\bibnamefont{Schweizer}},
  \bibnamefont{and} \bibinfo{author}{\bibfnamefont{L.~M.}
  \bibnamefont{Chayes}}, \bibinfo{journal}{Phys.\ Rev.\ E}
  \textbf{\bibinfo{volume}{54}}, \bibinfo{pages}{1332} (\bibinfo{year}{1996}).

\bibitem[{\citenamefont{Newman and Ziff}(2000)}]{Newman}
\bibinfo{author}{\bibfnamefont{M.~E.~J.} \bibnamefont{Newman}}
  \bibnamefont{and} \bibinfo{author}{\bibfnamefont{R.~M.} \bibnamefont{Ziff}},
  \bibinfo{journal}{Phys.\ Rev.\ Lett.} \textbf{\bibinfo{volume}{85}},
  \bibinfo{pages}{4104} (\bibinfo{year}{2000}).

\bibitem[{\citenamefont{Newman and Ziff}(2001)}]{NewmanPRE}
\bibinfo{author}{\bibfnamefont{M.~E.~J.} \bibnamefont{Newman}}
  \bibnamefont{and} \bibinfo{author}{\bibfnamefont{R.~M.} \bibnamefont{Ziff}},
  \bibinfo{journal}{Phys.\ Rev.\ E} \textbf{\bibinfo{volume}{64}},
  \bibinfo{pages}{016706} (\bibinfo{year}{2001}).

\bibitem[{\citenamefont{Hoshen and Kopelman}(1976)}]{Hoshen}
\bibinfo{author}{\bibfnamefont{J.}~\bibnamefont{Hoshen}} \bibnamefont{and}
  \bibinfo{author}{\bibfnamefont{R.}~\bibnamefont{Kopelman}},
  \bibinfo{journal}{Phys.\ Rev.\ B} \textbf{\bibinfo{volume}{14}},
  \bibinfo{pages}{3438} (\bibinfo{year}{1976}).

\bibitem[{\citenamefont{Hammersley and Handscomb}(1964)}]{Hammersley}
\bibinfo{author}{\bibfnamefont{J.~M.} \bibnamefont{Hammersley}}
  \bibnamefont{and} \bibinfo{author}{\bibfnamefont{D.~C.}
  \bibnamefont{Handscomb}}, \emph{\bibinfo{title}{Monte Carlo Methods}}
  (\bibinfo{publisher}{Methuen}, \bibinfo{address}{London},
  \bibinfo{year}{1964}).

\bibitem[{\citenamefont{Leath}(1976)}]{Leath}
\bibinfo{author}{\bibfnamefont{P.~L.} \bibnamefont{Leath}},
  \bibinfo{journal}{Phys.\ Rev.\ B} \textbf{\bibinfo{volume}{14}},
  \bibinfo{pages}{5046} (\bibinfo{year}{1976}).

\bibitem[{\citenamefont{Alexandrowicz}(1980)}]{Alexandrowicz}
\bibinfo{author}{\bibfnamefont{Z.}~\bibnamefont{Alexandrowicz}},
  \bibinfo{journal}{Phys.\ Lett.\ A} \textbf{\bibinfo{volume}{80}},
  \bibinfo{pages}{284} (\bibinfo{year}{1980}).

\bibitem[{\citenamefont{Tomita and Okabe}(2001)}]{Tomita}
\bibinfo{author}{\bibfnamefont{Y.}~\bibnamefont{Tomita}} \bibnamefont{and}
  \bibinfo{author}{\bibfnamefont{Y.}~\bibnamefont{Okabe}},
  \bibinfo{journal}{Phys.\ Rev.\ Lett.} \textbf{\bibinfo{volume}{86}},
  \bibinfo{pages}{572} (\bibinfo{year}{2001}).

\bibitem[{\citenamefont{Ziff}(1992)}]{Ziff}
\bibinfo{author}{\bibfnamefont{R.~M.} \bibnamefont{Ziff}},
  \bibinfo{journal}{Phys.\ Rev.\ Lett.} \textbf{\bibinfo{volume}{69}},
  \bibinfo{pages}{2670} (\bibinfo{year}{1992}).

\bibitem[{\citenamefont{Reynolds et~al.}(1980)\citenamefont{Reynolds, Stanley,
  and Klein}}]{Reynolds}
\bibinfo{author}{\bibfnamefont{P.~J.} \bibnamefont{Reynolds}},
  \bibinfo{author}{\bibfnamefont{H.~E.} \bibnamefont{Stanley}},
  \bibnamefont{and} \bibinfo{author}{\bibfnamefont{W.}~\bibnamefont{Klein}},
  \bibinfo{journal}{Phys.\ Rev.\ B} \textbf{\bibinfo{volume}{21}},
  \bibinfo{pages}{1223} (\bibinfo{year}{1980}).

\bibitem[{\citenamefont{Ziff and Newman}(2002)}]{ZiffPRE}
\bibinfo{author}{\bibfnamefont{R.~M.} \bibnamefont{Ziff}} \bibnamefont{and}
  \bibinfo{author}{\bibfnamefont{M.~E.~J.} \bibnamefont{Newman}},
  \bibinfo{journal}{Phys.\ Rev.\ E} \textbf{\bibinfo{volume}{66}},
  \bibinfo{pages}{016129} (\bibinfo{year}{2002}).

\bibitem[{\citenamefont{Knuth}(1981)}]{Knuth}
\bibinfo{author}{\bibfnamefont{D.~E.} \bibnamefont{Knuth}},
  \emph{\bibinfo{title}{The Art of Computer Programming, Volume 2:
  Seminumerical Algorithms}} (\bibinfo{publisher}{Addison-Wesley},
  \bibinfo{address}{Reading, MA}, \bibinfo{year}{1981}), \bibinfo{edition}{2nd}
  ed.

\bibitem[{\citenamefont{Matsumoto and Nishimura}(1998)}]{Matsumoto}
\bibinfo{author}{\bibfnamefont{M.}~\bibnamefont{Matsumoto}} \bibnamefont{and}
  \bibinfo{author}{\bibfnamefont{T.}~\bibnamefont{Nishimura}},
  \bibinfo{journal}{ACM Trans.\ Mod.\ Comp.\ Sim.}
  \textbf{\bibinfo{volume}{8}}, \bibinfo{pages}{3} (\bibinfo{year}{1998}).

\bibitem[{\citenamefont{Deng and Bl\"ote}(2005)}]{Deng}
\bibinfo{author}{\bibfnamefont{Y.}~\bibnamefont{Deng}} \bibnamefont{and}
  \bibinfo{author}{\bibfnamefont{H.~W.~J.} \bibnamefont{Bl\"ote}},
  \bibinfo{journal}{Phys.\ Rev.\ E} \textbf{\bibinfo{volume}{72}},
  \bibinfo{pages}{016126} (\bibinfo{year}{2005}).

\end{thebibliography}

\end{document}